# Isogeometric approach for nonlinear bending and post-buckling analysis of functionally graded plates under thermal environment

Loc V. Tran[1], Phuc Phung-Van[1], Jaehong Lee[2], H. Nguyen-Xuan[2,3*], M. Abdel Wahab[1]

[1] Department of Mechanical Construction and Production, Faculty of Engineering and Architecture, Ghent University, 9000, Ghent – Belgium

[2] Department of Architectural Engineering, Sejong University, 98 Kunja Dong, Kwangjin Ku, Seoul, 143-747, South Korea

[3] Center for Interdisciplinary Research in Technology (CIRTech), Ho Chi Minh City University of Technology (HUTECH), Ho Chi Minh City 700000, Vietnam

**Abstract**

In this paper, equilibrium and stability equations of functionally graded material (FGM) plate under thermal environment are formulated based on isogeometric analysis (IGA) in combination with higher-order shear deformation theory (HSDT). The FGM plate is made by a mixture of two distinct components, for which material properties not only vary continuously through thickness according to a power-law distribution but also are assumed to be a function of temperature. Temperature field is assumed to be constant in any plane and uniform, linear and nonlinear through plate thickness, respectively. The governing equation is in nonlinear form based on von Karman assumption and thermal effect. A NURBS-based isogeometric finite element formulation is utilized to naturally fulfil the rigorous $C^1$-continuity required by the present plate model. Influences of gradient indices, boundary conditions, temperature distributions, material properties, length-to-thickness ratios on the behaviour of FGM plate are discussed in details. Numerical results demonstrate excellent performance of the present approach.

---

[*] Corresponding author

Email address: tranvinhloc@gmail.com (Loc V. Tran), ngx.hung@hutech.edu.vn (H. Nguyen-Xuan), Magd.AbdelWahab@ugent.be (M. Abdel Wahab)

# 1    Introduction

Laminated composite plates made by stacking several lamina layers together possess many favourable mechanical properties, e.g. wear resistance, high ratio of stiffness, strength-to-weight ratios, etc. Therefore, they are extensively used in aerospace, aircraft structures, high-speed vehicle frames, etc. However, an important feature in their designs is thermal effect. For an example, the space vehicles flying at hypersonic speeds experience extremely rapid temperature rise in very short time from aerodynamic heating due to friction between the vehicle surface and the atmosphere, i.e. in U.S. space shuttles, the temperature on their outside surface increases to an attitude of 1500°C for a few minutes [1]. This can lead to harmful effects due to stress concentration, cracking and de-bonding, which can occur at the interface between two distinct layers [2, 3]. To overcome this shortcoming, a group of scientists in Sendai-Japan proposed an advanced material, so–called functionally graded materials (FGMs) [4-6]. The most common FGMs are the mixtures of a ceramic and a metal, for which material properties vary smoothly and continuously in a predetermined direction. Consequently, they enable to reduce the thermal stresses due to smoothly transitioning the properties of the components. Furthermore, they inherit the best properties of the distinct components, e.g. low thermal conductivity, high thermal resistance by ceramic part, ductility, durability and superiority of fracture toughness of metal part. FGMs are now developed as the structural components in many engineering applications [1].

In order to clearly understand the scientific and engineering communities in the field of modelling, analysis and design of FGM plate structures, many studies have been reported by various researchers. For instant, Praveen and Reddy [7] studied the nonlinear transient responses of FGM plates under thermal and mechanical loadings using FEM with von Karman assumptions. Vel and Batra [8, 9] obtained the three dimensional exact solutions for the thermo-elastic deformation of FGM rectangular plate. Javaheri and Eslami [10, 11] investigated thermal buckling behaviour of the FGM plates. Ferreira et al. [12, 13] performed static and dynamic analysis of FGM plate based on HSDT using the mesh-free method. Park and Kim [14] investigated thermal post buckling and vibration analyses of simply supported FGM plates by using FEM. Lee et al. [15, 16] developed the element-free *kp*-Ritz method to study behaviour of FGM plate. Nguyen-Xuan et al. [17-19] developed smoothed finite element method based on triangular element to study thermo-elastic bending, free vibration and elastic stability of FGM plates and so on.

In the aforementioned studies, it can be seen that for modelling the plate structures, the formulation may be reduced to a linear problem based on small displacement and strain assumptions. Linear solution can be obtained easily with low computational cost and sometime is a reasonable idealization. However, linear solution usually deviates from real response of structures [20-23]. In some cases, assumption of nonlinearity is the only option for analyst, i.e. post buckling phenomenon [24, 25]. In that case the structures experience large deformation. So, geometrically nonlinear analysis is employed to fully investigate the plate behaviour in the large deformation regime. Furthermore, several plate theories are provided to predict accurately the structure responses. Among them, classical plate theory (CPT) requires $C^1$-continuity elements and merely provides acceptable results for thin plate, whilst first order shear deformation theory (FSDT) is suitable for moderate and thick plate. However, it describes incorrect shear energy part. And the standard FSDT-based finite elements are too stiff and lead to shear locking [26, 27]. Therefore, HSDT models, which take into account higher-order variations of the in-plane displacements through thickness, are proposed. Consequently, they enable to really describe shear strain/stress distributions with the non-linear paths and traction-free boundary condition at the top and bottom surfaces of the plate. Moreover, the HSDT models provide better results and yield more accurate and stable solutions. However, the HSDT requires $C^1$-continuity elements that cause the obstacles in the standard finite element formulations. Fortunately, Hughes and his co-worker proposed a novel numerical method – so-called isogeometric analysis (IGA) [28, 29], which yields higher-order continuity naturally and easily. The core idea of this method is to integrate both geometric description and finite element approximation through the same basis function space of B-spline or NURBS. The major strengths of this method are that it is flexible to control the high continuity of basis shape functions, e.g., $C^{p-1}$-continuity for $p^{\text{th}}$-order NURBS. Furthermore, by removal mesh generation feature, this method produces a seamless integration of computer aid design (CAD) and finite element analysis (FEA) tools. As a result, IGA simplifies the cost-intensive computational model generation procedure which is the major bottleneck in engineering analysis-design [30]. After ten years of development, IGA has been widely applied in engineering and among them with FGM plate structures. For example, Valizadeh et al. [31] and Yin et al. [32] employed this method to study the static and dynamic behaviours of FGM plates based on FSDT. Tran et al. [26, 27, 33] studied the static bending, buckling load and also natural frequency of intact FGM plates and cracked ones [34] based on HSDT and then extended their previous work for thermal buckling analysis with various types of temperature distribution [35]. Recently, the geometrically nonlinear problems in FGM plates based on von Karman assumptions are reported by Yin et al. [36] and Jari et al. [37] based on FSDT. Apparently, there are no researches on developing HSDT model, which can exploit well the high continuity of NURBS, in nonlinear bending analysis and especially in thermal post-buckling analysis of FGM plates. To fill this research gap, we investigate an efficient computational approach based on IGA and HSDT for behaviour of FGM plates in thermal environment.

The paper is outlined as follows. The next section introduces the theoretical formulation for functionally graded plate. The von Karman assumption is employed to depict behaviour of the plate structure in the large deformation regime. Assumption of temperature field due to uniform, linear and nonlinear distribution through the plate thickness is described in section 3. Section 4 presents a framework of isogeometric analysis for the plate structure. Section 5 gives the solution procedure for the plate problems which can be categorized into two groups: geometrically nonlinear and nonlinear-eigenvalue analysis for tracing the post-buckling paths. The present formulation is verified firstly by comparing with other available results in the literature and the influences of gradient indices, boundary conditions, temperature distributions, material properties and length-to-thickness ratio on the behaviour of FGM plate are then examined in section 6. Finally, this article is closed with some concluding remarks.

## 2 A background on functionally graded plates

### *2.1 Functionally graded material*

Functionally graded material is a composite material, which is commonly fabricated by mixing two distinct material phases, i.e. ceramic and metal, for which properties change continuously along certain dimensions of the structure, as shown in Figure 1. It is assumed that the volume fractions of the material phases are given by the power-law type function and satisfy the unity, i.e.

$$V_c(z) = \left(\frac{1}{2} + \frac{z}{h}\right)^n \quad , \quad V_c + V_m = 1 \tag{1}$$

where $n \in \mathbb{R}^+$ is the power index or gradient index Then, the effective material properties such as the Young's modulus ($E$), shear modulus ($\mu$), Poisson's ratio ($\nu$), the density ($\rho$), thermal conductivity ($k$) and thermal expansion ($\alpha$) can be estimated according to the rule of mixture as follows

$$P_e = P_c V_c + P_m V_m \tag{2}$$

As noted that the subscripts *m*, *c* and *e* refer to metal, ceramic and effective constituents, respectively.

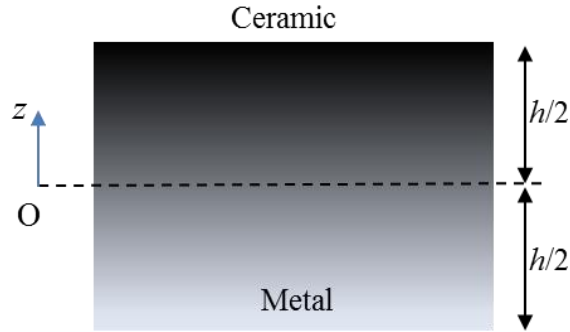

*Figure 1 A Functionally graded material layer.*

Figure 2 illustrates the distribution of the effective Young's modulus through thickness of Al/Al$_2$O$_3$ FGM plate via the power index *n*. As observed, $n = 0 \rightarrow V_c = 1, V_m = 0$, the structure is fully ceramic and when $n = \infty \rightarrow V_c = 0, V_m = 1$, the homogeneous metal is retrieved. Moreover, $V_c(h/2)=1$ and $V_m(-h/2)=1$ means that fully ceramic and metal phase on the top and the bottom surfaces, respectively.

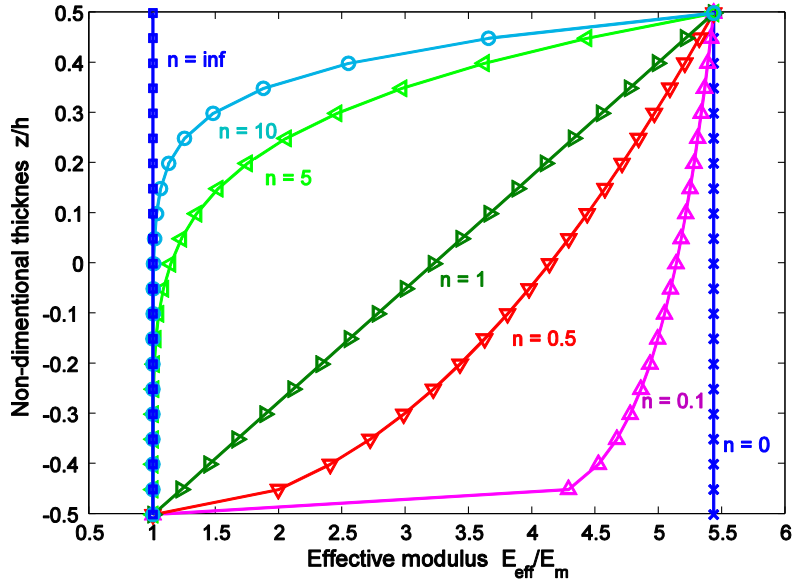

*Figure 2 The effective modulus of Al/Al$_2$O$_3$ FGM plate.*

In thermal environment, high temperature makes a significant change in mechanical properties of the constituent materials. Therefore, it is essential to take into account the temperature-dependent material property to accurately predict the mechanical responses of FGM structures. According to Ref. [38], the properties of the common structural ceramics and metals are expressed as a nonlinear function of temperature

$$P = P_0 \left( P_{-1} T^{-1} + 1 + P_1 T + P_2 T^2 + P_3 T^3 \right) \tag{3}$$

where $P_0$, $P_{-1}$, $P_1$, $P_2$ and $P_3$ are the coefficients of temperature, which can be found in Ref. [39] as unique parameters for each constituent material.

### *2.2 Plate formulation*

According to the generalized shear deformation plate theory [27], the displacement of an arbitrary point $\mathbf{u} = \{u, v, w\}^T$ can be written as

$$\mathbf{u} = \mathbf{u}_1 + z\mathbf{u}_2 + f(z)\mathbf{u}_3 \tag{4}$$

where $\mathbf{u}_1 = \{u_0 \ v_0 \ w_0\}^T$ is the displacement components in $x$, $y$ and $z$ axes, $\mathbf{u}_2 = -\{w_{0,x} \ w_{0,y} \ 0\}^T$ and $\mathbf{u}_3 = \{\beta_x \ \beta_y \ 0\}^T$ are the rotations in the $xz$, $yz$ and $xy$ planes, respectively. The distributed function is choose following to Reddy's theory [40] as $f(z) = z - 4z^3/(3h^2)$.

Enforcing the assumptions of small strains, moderate rotations and large displacements, the von Karman nonlinear theory is adopted in strain-displacement relations as follows [41]

$$\begin{Bmatrix} \varepsilon_x \\ \varepsilon_y \\ \gamma_{xy} \\ \gamma_{xz} \\ \gamma_{yz} \end{Bmatrix} = \begin{Bmatrix} u_{,x} \\ v_{,y} \\ u_{,y} + v_{,x} \\ u_{,z} + w_{,x} \\ v_{,z} + w_{,y} \end{Bmatrix} + \frac{1}{2}\begin{Bmatrix} w_{,x}^2 \\ w_{,y}^2 \\ 2w_{,x}w_{,y} \\ 0 \\ 0 \end{Bmatrix} \tag{5}$$

As using the assumed displacement field in Eq. (4), the strain vector with separated in-plane strain $\boldsymbol{\varepsilon}$ and shear strain $\boldsymbol{\gamma}$ are given as

$$\begin{Bmatrix} \boldsymbol{\varepsilon} \\ \boldsymbol{\gamma} \end{Bmatrix} = \begin{Bmatrix} \boldsymbol{\varepsilon}_m \\ 0 \end{Bmatrix} + \begin{Bmatrix} z\boldsymbol{\kappa}_1 \\ 0 \end{Bmatrix} + \begin{Bmatrix} f(z)\boldsymbol{\kappa}_2 \\ f'(z)\boldsymbol{\beta} \end{Bmatrix} \tag{6}$$

where the in-plane, the bending and the shear strains are defined, respectively,

$$\boldsymbol{\varepsilon}_m = \begin{bmatrix} u_{0,x} \\ v_{0,y} \\ u_{0,y} + v_{0,x} \end{bmatrix} + \frac{1}{2}\begin{bmatrix} w_{0,x}^2 \\ w_{0,x}^2 \\ 2w_{0,x}w_{0,y} \end{bmatrix} = \boldsymbol{\varepsilon}_L + \boldsymbol{\varepsilon}_{NL}$$

$$\boldsymbol{\kappa}_1 = -\begin{bmatrix} w_{0,xx} \\ w_{0,yy} \\ 2w_{0,xy} \end{bmatrix}, \quad \boldsymbol{\kappa}_2 = \begin{bmatrix} \beta_{x,x} \\ \beta_{y,y} \\ \beta_{x,y} + \beta_{y,x} \end{bmatrix}, \quad \boldsymbol{\beta} = \begin{bmatrix} \beta_x \\ \beta_y \end{bmatrix} \tag{7}$$

In Eq. (7) the nonlinear component of in-plane strain can be rewritten as

$$\varepsilon_{NL} = \frac{1}{2}\mathbf{A}_\theta \boldsymbol{\theta} \tag{8}$$

where

$$\mathbf{A}_\theta = \begin{bmatrix} w_{0,x} & 0 \\ 0 & w_{0,y} \\ w_{0,y} & w_{0,x} \end{bmatrix} \quad \text{and} \quad \boldsymbol{\theta} = \begin{Bmatrix} w_{0,x} \\ w_{0,y} \end{Bmatrix} \tag{9}$$

Considering thermal effect, the thermal strain is given by

$$\boldsymbol{\varepsilon}^{th} = \alpha_e(z)\Delta T(z)\begin{bmatrix} 1 & 1 & 0 \end{bmatrix}^T \tag{10}$$

in which $\alpha_e(z)$ is the effective thermal coefficient according to Eq. (2) and $\Delta T$ is the temperature change defined as

$$\Delta T(z) = T(z) - T_i \tag{11}$$

where $T_i$ is the initial temperature and $T(z)$ is the current temperature.

In these plate theories, the transverse normal stress $\sigma_z$ is assumed to be zero. Hence, the reduced constitutive relation for the FGM plate is given by

$$\begin{Bmatrix} \boldsymbol{\sigma} \\ \boldsymbol{\tau} \end{Bmatrix} = \begin{bmatrix} \mathbf{C} & \mathbf{0} \\ \mathbf{0} & \mathbf{G} \end{bmatrix} \begin{Bmatrix} \boldsymbol{\varepsilon} - \boldsymbol{\varepsilon}^{th} \\ \boldsymbol{\gamma} \end{Bmatrix} \tag{12}$$

where the material matrices are given as

$$\mathbf{C} = \frac{E_e}{1-\nu_e^2}\begin{bmatrix} 1 & \nu_e & 0 \\ \nu_e & 1 & 0 \\ 0 & 0 & (1-\nu_e)/2 \end{bmatrix} \tag{13}$$

$$\mathbf{G} = \frac{E_e}{2(1+\nu_e)}\begin{bmatrix} 1 & 0 \\ 0 & 1 \end{bmatrix} \tag{14}$$

The in-plane forces, moments and shear forces are calculated by

$$\begin{Bmatrix} \mathbf{N} \\ \mathbf{M} \\ \mathbf{P} \end{Bmatrix} = \int_{-h/2}^{h/2} \boldsymbol{\sigma} \begin{Bmatrix} 1 \\ z \\ f(z) \end{Bmatrix} dz \quad \text{and} \quad \mathbf{Q} = \int_{-h/2}^{h/2} f'(z) \boldsymbol{\tau} dz \tag{15}$$

Substituting Eq. (12) into Eq.(15), stress resultants are rewritten in matrix form as

$$\underbrace{\begin{Bmatrix} \mathbf{N} \\ \mathbf{M} \\ \mathbf{P} \\ \mathbf{Q} \end{Bmatrix}}_{\hat{\boldsymbol{\sigma}}} = \underbrace{\begin{bmatrix} \mathbf{A} & \mathbf{B} & \mathbf{E} & \mathbf{0} \\ \mathbf{B} & \mathbf{D} & \mathbf{F} & \mathbf{0} \\ \mathbf{E} & \mathbf{F} & \mathbf{H} & \mathbf{0} \\ \mathbf{0} & \mathbf{0} & \mathbf{0} & \mathbf{D}^S \end{bmatrix}}_{\hat{\mathbf{D}}} \underbrace{\begin{Bmatrix} \boldsymbol{\varepsilon}_m \\ \boldsymbol{\kappa}_1 \\ \boldsymbol{\kappa}_2 \\ \boldsymbol{\beta} \end{Bmatrix}}_{\hat{\boldsymbol{\varepsilon}}} - \underbrace{\begin{Bmatrix} \mathbf{N}^{th} \\ \mathbf{M}^{th} \\ \mathbf{P}^{th} \\ 0 \end{Bmatrix}}_{\hat{\boldsymbol{\sigma}}_0} \tag{16}$$

in which

$$A_{ij}, B_{ij}, D_{ij}, E_{ij}, F_{ij}, H_{ij} = \int_{-h/2}^{h/2} (1, z, z^2, f(z), zf(z), f^2(z)) C_{ij} dz$$

$$D_{ij}^s = \int_{-h/2}^{h/2} [f'(z)]^2 G_{ij} dz \tag{17}$$

and the thermal stress resultants are the functions of the incremental temperature $\Delta T$

$$\{\mathbf{N}^{th} \quad \mathbf{M}^{th} \quad \mathbf{P}^{th}\} = \int_{-h/2}^{h/2} \mathbf{C} \begin{Bmatrix} \alpha_e \\ \alpha_e \\ 0 \end{Bmatrix} \{1 \quad z \quad z^3\} \Delta T dz \tag{18}$$

It is evident that the function $f'(z) = 1 - 4z^2/h^2$ is a parabolic function of thickness and produces zero values at $z = \pm h/2$. It means that the traction-free boundary condition is automatically satisfied at the top and bottom plate surfaces. Furthermore, the transverse shear forces are described parabolically through the plate thickness. Hence, the shear correction factors are no longer required in this model.

Employing the principle of virtual displacement, the variation of total energy of the plate can be derived by

$$\delta \Pi = \delta U_\varepsilon - \delta V = \int_\Omega \delta \hat{\boldsymbol{\varepsilon}}^T \hat{\boldsymbol{\sigma}} d\Omega - \int_\Omega \delta \mathbf{u}^T f_z d\Omega = 0 \tag{19}$$

where $f_z$ is the transverse load.

## 3 Type of temperature distribution

Under thermal environment, the temperature is assumed to be uniform on the top and bottom surfaces and varies through the plate thickness. Some case studies are given as

### 3.1 *Uniform temperature rise*

It is assumed that the reference temperature initially equals to $T_i$ and then uniformly increases to a final value at which the plate is bucked. Therefore, the temperature change $\Delta T = T_f - T_i$ is constant everywhere in the plate. Substituting it into Eq. (18) leads to the critical buckling temperature as follows:

$$\Delta T_{cr} = N_{cr}^{th} / \tilde{X}$$

where
$$\tilde{X} = \int_{-h/2}^{h/2} \frac{E_e(z)}{1+\nu_e(z)} \alpha_e(z) \mathrm{d}z$$
(20)

### 3.2 *Linear temperature across the plate thickness*

Considering a FGM plate, which initial temperature at the ceramic-rich and metal-rich surfaces are $T_c$ and $T_m$, respectively. Temperature is assumed to be linear distribution through the plate thickness by

$$T(z) = (T_c - T_m)\left(\frac{z}{h} + \frac{1}{2}\right) + T_m,$$
(21)

Substituting Eq. (21) into Eq. (11) and then solving Eq. (18), the critical buckling temperature difference between two plate surfaces $\Delta T = T_c - T_m$ is calculated as

$$\Delta T_{cr} = \frac{N_{cr}^{th} - \tilde{X}(T_m - T_i)}{\tilde{Y}}$$
(22)

where
$$\tilde{Y} = \int_{-h/2}^{h/2} \frac{E_e(z)\alpha_e(z)}{1+\nu_e(z)} \left(\frac{z}{h}+\frac{1}{2}\right) dz$$

### 3.3 Non-linear temperature change across the thickness

The temperature field in the FGM plate follows the one-dimensional steady state heat conduction equation and the boundary conditions are given by

$$-\frac{d}{dz}\left(k(z)\frac{dT}{dz}\right) = 0 \quad , T(h/2) = T_c \quad , T(-h/2) = T_m \quad (23)$$

The solution of Eq. (23) is obtained in Fourier series [42, 43] as:

$$T(z) = T_m + \eta(z)(T_c - T_m) \quad (24)$$

where

$$\eta(z) = \left(\frac{z}{h}+\frac{1}{2}\right)\sum_{i=0}^{\infty}\frac{1}{ni+1}\left(\frac{z}{h}+\frac{1}{2}\right)^{ni}\left(\frac{k_m - k_c}{k_m}\right)^i \Big/ \sum_{i=0}^{\infty}\frac{1}{ni+1}\left(\frac{k_m - k_c}{k_m}\right)^i \quad (25)$$

Figure 3 illustrates the effect of the gradient index *n* on the temperature distribution through the thickness of the Al/Al$_2$O$_3$ FGM plate subjected to a thermal load where the top and bottom surfaces are held at 300°C and 20°C, respectively. It is evident that the temperature in the FGM plates follows a nonlinear distribution and is always lower than that in the homogenous plates. In addition, there is linearly distributed temperature through thickness as same as Eq. (21) in case of the homogeneous plate.

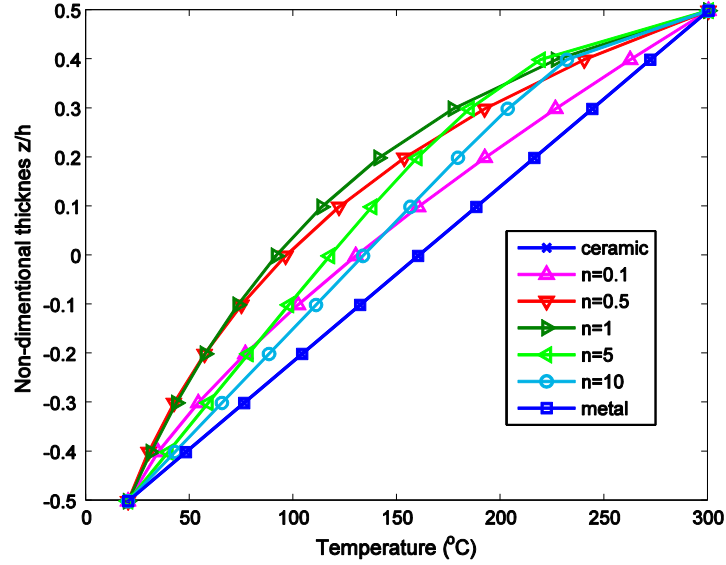

*Figure 3* Temperature distributions through the thickness of Al/Al$_2$O$_3$ FGM plate.

Being similar to the previous types, after solving Eq. (18) with temperature field described in Eq. (24), the critical buckling temperature difference between two opposite plate surfaces becomes

$$\Delta T_{cr} = \frac{N_{cr}^{th} - \tilde{X}(T_m - T_i)}{\tilde{Z}}$$

(26)

where $\tilde{Z} = \int_{-h/2}^{h/2} \frac{E_e(z)\alpha_e(z)}{1+\nu_e(z)} \eta(z) \mathrm{d}z$

## 4  On isogeometric nonlinear analysis of plate structure

### 4.1  *A brief of isogeometric analysis*

Isogeometric approach (IGA) is proposed by Hughes and his co-workers [28] with the primary original purpose is to enable a tighter connection between computer aided design (CAD) and finite element analysis (FEA). The main idea of this method is to utilize the same basis functions such as: B-spline, non-uniform rational B-spline (NURBS), etc. in both geometry description and finite approximation. A B-splines basis of degree *p* is generated from a non-decreasing sequence of parameter value $\xi_i$, $i = 1,...n+p$, called a knot vector $\Xi = \{\xi_1, \xi_2, ..., \xi_{n+p+1}\}$, in which $\xi_1 \leq \xi_2 \leq ... \leq \xi_{n+p+1}$. $\xi_i \in \mathbb{R}$ is the $i^{th}$ knot and $n$ is number of the basis functions. In the so-called open knot, the first and the last knots are repeated by $p+1$ times and very often get values of $\xi_1 = 0$ and $\xi_{n+p+1} = 1$.

Using Cox-de Boor algorithm, the univariate B-spline basis functions $N_{i,p}(\xi)$ are defined recursively on the corresponding knot vector

$$N_i^p(\xi) = \frac{\xi - \xi_i}{\xi_{i+p} - \xi_i} N_i^{p-1}(\xi) + \frac{\xi_{i+p+1} - \xi}{\xi_{i+p+1} - \xi_{i+1}} N_{i+1}^{p-1}(\xi)$$

$$\text{as } p = 0, \quad N_i^0(\xi) = \begin{cases} 1 & \text{if } \xi_i \leq \xi < \xi_{i+1} \\ 0 & \text{otherwise} \end{cases}$$

(27)

By a simple way, so-called tensor product of univariate B-splines, the multivariate B-spline basis functions are generated

$$N_i^p(\boldsymbol{\xi}) = \prod_{\alpha=1}^{d} N_{i_\alpha}^{p_\alpha}(\boldsymbol{\xi}^\alpha) \tag{28}$$

where parametric $d = 1, 2, 3$ according to 1D, 2D and 3D spaces, respectively. Figure 4 gives an illustration of bivariate B-splines basic based on tensor product of two knot vectors $\Xi = \{0, 0, 0, \frac{1}{5}, \frac{2}{5}, \frac{3}{5}, \frac{3}{5}, \frac{4}{5}, 1, 1, 1\}$ and $\mathbf{H} = \{0, 0, 0, 0, \frac{1}{3}, \frac{1}{3}, \frac{2}{3}, \frac{2}{3}, 1, 1, 1, 1\}$ in two parametric dimensions $\xi$ and $\eta$, respectively.

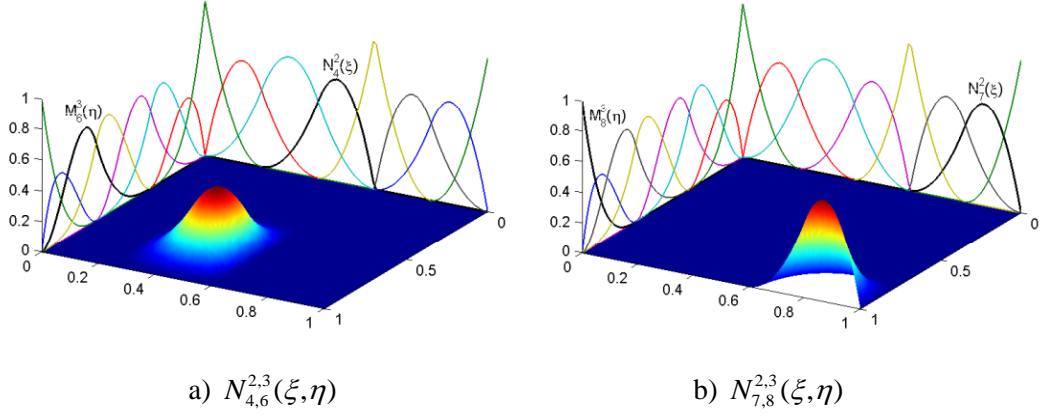

a) $N_{4,6}^{2,3}(\xi,\eta)$  b) $N_{7,8}^{2,3}(\xi,\eta)$

*Figure 4 Bivariate B-splines basic functions.*

After defining the B-spline basis functions, a domain, including B-spline curve, surface or solid, can be constructed from a linear combination of them with control points $\mathbf{P}_i$

$$\mathbf{S}(\xi) = \sum_i N_i^p(\xi) \mathbf{P}_i \tag{29}$$

However, for some conic shapes (e.g., circles, ellipses, spheres, etc.), NURBS offer a more generalized way in form of rational functions as

$$R_i^p(\xi) = N_i^p(\xi)\zeta_i \Big/ \sum_j N_j^p(\xi)\zeta_j \tag{30}$$

where $\zeta_i > 0$ is the so-called individual weight corresponding to B-splines basis functions $N_i^p(\xi)$. It is seen that NURBS basic will become B-spline, when the individual weight is constant.

### 4.2  Discrete system equation

Being different from traditional finite element method, which utilizes the Lagrange basis functions in approximating the unknown solutions and the geometry, NURBS-based IGA employs the NURBS basis ones from geometric description to construct the approximated solution

$$\mathbf{u}^h(\xi) = \sum_A R_A(\xi)\mathbf{q}_A \tag{31}$$

where $\mathbf{q}_A = \begin{bmatrix} u_{0A} & v_{0A} & \beta_{xA} & \beta_{yA} & w_{0A} \end{bmatrix}^T$ denotes the vector of nodal degrees of freedom associated with the control point $\mathbf{P}_A$.

Substituting Eq.(31) into Eq. (7), the generalized strains can be rewritten in matrix form as:

$$\hat{\boldsymbol{\varepsilon}} = \left( \mathbf{B}^L + \frac{1}{2} \mathbf{B}^{NL} \right) \mathbf{q}$$

where $\mathbf{B}^L$ is the linear infinitesimal strain

$$\mathbf{B}_A^L = \left[ \left( \mathbf{B}_A^m \right)^T \ \left( \mathbf{B}_A^{b1} \right)^T \ \left( \mathbf{B}_A^{b2} \right)^T \ \left( \mathbf{B}_A^s \right)^T \right]^T \tag{32}$$

in which

$$\mathbf{B}_A^m = \begin{bmatrix} R_{A,x} & 0 & 0 & 0 & 0 \\ 0 & R_{A,y} & 0 & 0 & 0 \\ R_{A,y} & R_{A,x} & 0 & 0 & 0 \end{bmatrix},$$

$$\mathbf{B}_A^{b1} = -\begin{bmatrix} 0 & 0 & R_{A,xx} & 0 & 0 \\ 0 & 0 & R_{A,yy} & 0 & 0 \\ 0 & 0 & 2R_{A,xy} & 0 & 0 \end{bmatrix}, \quad \mathbf{B}_A^{b2} = \begin{bmatrix} 0 & 0 & 0 & R_{A,x} & 0 \\ 0 & 0 & 0 & 0 & R_{A,y} \\ 0 & 0 & 0 & R_{A,y} & R_{A,x} \end{bmatrix},$$

$$\mathbf{B}_A^s = \begin{bmatrix} 0 & 0 & 0 & R_A & 0 \\ 0 & 0 & 0 & 0 & R_A \end{bmatrix}, \quad \mathbf{B}_A^g = \begin{bmatrix} 0 & 0 & R_{A,x} & 0 & 0 \\ 0 & 0 & R_{A,y} & 0 & 0 \end{bmatrix}$$

and the nonlinear strain matrix $\mathbf{B}^{NL}$ is found to be a linear function of the displacement

$$\mathbf{B}_A^{NL}(\mathbf{q}) = \begin{bmatrix} \mathbf{A}_\theta \\ \mathbf{0} \end{bmatrix} \mathbf{B}_A^g \tag{33}$$

Variation of the strain is defined as

$$\delta \hat{\boldsymbol{\varepsilon}} = \left( \mathbf{B}^L + \mathbf{B}^{NL} \right) \delta \mathbf{q} \tag{34}$$

Substituting Eqs. (16) and (34) into Eq. (19) and eliminating the virtual displacement vector $\delta \mathbf{q}^T$, the governing equation can be written in the following matrix form

$$\left(\mathbf{K}_L + \mathbf{K}_{NL} - \mathbf{K}_0\right)\mathbf{q} = \mathbf{F} \tag{35}$$

in which $\mathbf{K}_L$ and $\mathbf{K}_{NL}$ are the linear and nonlinear stiffness matrices, respectively, whilst $\mathbf{K}_0$ is the initial stress stiffness matrix due to the initial compressive load by temperature

$$\mathbf{K}_L = \int_\Omega \left(\mathbf{B}^L\right)^T \hat{\mathbf{D}} \mathbf{B}^L d\Omega \tag{36}$$

$$\mathbf{K}_{NL} = \frac{1}{2}\int_\Omega \left(\mathbf{B}^L\right)^T \hat{\mathbf{D}} \mathbf{B}^{NL} d\Omega + \int_\Omega \left(\mathbf{B}^{NL}\right)^T \hat{\mathbf{D}} \mathbf{B}^L d\Omega + \int_\Omega \frac{1}{2}\left(\mathbf{B}^{NL}\right)^T \hat{\mathbf{D}} \mathbf{B}^{NL} d\Omega \tag{37}$$

$$\mathbf{K}_0 = \int_\Omega \left(\mathbf{B}^g\right)^T \begin{bmatrix} N_x^{th} & N_{xy}^{th} \\ N_{xy}^{th} & N_y^{th} \end{bmatrix} \mathbf{B}^g d\Omega \tag{38}$$

and $\mathbf{F}$ is the load vector depending on mechanical and thermal loads

$$\mathbf{F} = \int_\Omega \left(\mathbf{B}^L\right)^T \hat{\boldsymbol{\sigma}}_0 + \mathbf{R}^T f_z d\Omega \tag{39}$$

## 5 Solution procedure

Depending on value of load vector, nonlinear analysis of FGM plates can be classified into two groups: geometrical nonlinear analysis and nonlinear eigenvalue analysis.

### 5.1 Geometrical nonlinear analysis

To solve the nonlinear equilibrium equation in Eq. (35), an iterative Newton-Raphson technique is employed. Let introduce a residual force as

$$\boldsymbol{\varphi}(\mathbf{q}) = \left(\mathbf{K}_L + \mathbf{K}_{NL}(\mathbf{q}) - \mathbf{K}_0\right)\mathbf{q} - \mathbf{F}^{ext} \to 0 \tag{40}$$

The residual force represents the error in this approximation and tends to zero during iteration. If $^i\mathbf{q}$, the approximate trial solution at the $i^{th}$ iteration, makes unbalance residual force an improved solution $^{i+1}\mathbf{q}$ is then suggested as

$$^{i+1}\mathbf{q} = {^i}\mathbf{q} + \Delta\mathbf{q} \tag{41}$$

The increment displacement can be defined by

$$\Delta\mathbf{q} = \left[\mathbf{F} - \left(\mathbf{K}_L + {^i}\mathbf{K}_{NL}({^i}\mathbf{q}) - \mathbf{K}_0\right){^i}\mathbf{q}\right]/\mathbf{K}_T \tag{42}$$

where $\mathbf{K}_T$ is called tangent stiffness matrix is defined as

$$\mathbf{K}_T = \frac{\partial \boldsymbol{\varphi}({^i}\mathbf{q})}{\partial\, {^i}\mathbf{q}} = \tilde{\mathbf{K}}_{NL} + \mathbf{K}_g \tag{43}$$

in which the matrix $\tilde{\mathbf{K}}_{NL}$ is strongly dependent on displacement

$$\tilde{\mathbf{K}}_{NL} = \int_\Omega \left(\mathbf{B}^L + \mathbf{B}^{NL}\right)^T \hat{\mathbf{D}} \left(\mathbf{B}^L + \mathbf{B}^{NL}\right) \mathrm{d}\Omega \tag{44}$$

and the geometric stiffness matrix is given by

$$\mathbf{K}_g = \int_\Omega \left(\mathbf{B}^g\right)^T \begin{bmatrix} N_x & N_{xy} \\ N_{xy} & N_y \end{bmatrix} \left(\mathbf{B}^g\right) \mathrm{d}\Omega \tag{45}$$

It is noted that being different from the initial stress stiffness matrix, $\mathbf{K}_0$, the geometric stiffness matrix is calculated due to the internal forces according to Eq. (16).

## 5.2 Nonlinear eigenvalue analysis

For a case of the homogeneous plates, under uniform temperature rise the thermal moments in Eq. (18) are equal to zero and only membrane forces are generated. Thus, the initially perfect plate is still flat with no transverse deflection. As a result, there is no effect of geometrical nonlinearity and Eq. (35) is simplified as

$$\left(\mathbf{K}_L - \lambda \mathbf{K}_0\right)\mathbf{q} = \mathbf{0} \tag{46}$$

where $\lambda \in \mathbb{R}^+$ is the load factor.

This is called linear buckling equation in order to determine the critical value of loading for a particular plate. As temperature increases to a critical point, the plate suddenly bucks and may lose its load carrying capacity but it is typically capable of working and carrying considerable additional load before the collapse or ultimate load is reached. In some cases this is even several times higher than the critical load [44]. This is called the post-buckling phenomenon. At this time, the plate structure undergoes a large deformation. Therefore, the effect of geometric nonlinearity based on von Karman nonlinear strain must be consider in governing equation as:

$$\left(\mathbf{K}_L + \mathbf{K}_{NL} - \lambda \mathbf{K}_0\right)\mathbf{q} = 0 \qquad (47)$$

In case of FGM plate, because of un-symmetric material distribution through the thickness, bending moments, which forces the plate laterally deform, develop together with the membrane forces during temperature change. Consequently, the plate is deflected as soon as thermal load is applied. Thus, the bifurcation phenomenon does not occur. However, for a special case, that is clamped edges, the supports are capable of handling the produced thermal moments [25, 45-47]. It maintains the plate in un-deformed pre-buckling state. Therefore, buckling bifurcation phenomenon does exist. FGM is also a function of temperature as shown in Eq.(3). Thus, solution of Eq. (47), which is a function of both the nodal variables $\mathbf{q}$ and temperature $T(z)$, should be solved by the incremental iterative methodology.

Firstly, using thermo-elastic properties at $T_m$ (the final temperature at the plate bottom), the smallest eigenvalue (load factor) and its corresponding eigenvector are obtained from the linear eigenvalue equation, Eq. (46). The buckling load, computed from multiplying the initial load with the load factor, is utilized to calculate the critical buckling temperature difference using Eqs. (20), (22) and (26) according to the type of temperature distribution. Next, the thermo-elastic properties at $T = T_m + \Delta T_{cr}$ is updated. Besides, the eigenvector is normalized and scaled up to desired amplitude to make sure that its magnitude is kept constant for each displacement incremental step. Then it is used as the displacement vector for evaluation of the nonlinear stiffness. Equation (47) is solved to obtain the load factor and the associated eigenvector. Subsequently, updated temperature $T$ is implemented. Convergence is verified by using a desired tolerance, i.e. $\varepsilon = 0.01$. If this is not satisfied, all the matrices are updated at the updated temperature by current load factor and displacement vector according to current buckling mode shape. Equation (47) is solved again to obtain the load factor and buckling mode shape. This iterative procedure keeps going until the convergence of the thermal buckling temperature is achieved.

## 6  Numerical examples

This section focuses on studying the nonlinear behaviour of FGM plate, which material properties are listed in Table 1, under transverse and thermal load. It is assumed in the latter that the temperature is uniform on the top and bottom surfaces and varies through the thickness direction as a constant, linear or nonlinear function. In these problems, we assume that the plate is constrained on all edges by:

- Simply supported condition, which is divided in two cases: movable and immovable in the in-plane directions.

$$\text{Movable edge (SSSS1):} \quad \begin{cases} v_0 = w_0 = \beta_y = 0 & \text{on } x = 0,\ L \\ u_0 = w_0 = \beta_x = 0 & \text{on } y = 0,\ W \end{cases} \tag{48}$$

$$\text{Immovable edge (SSSS2):} \quad \begin{cases} u_0 = v_0 = w_0 = \beta_y = 0 & \text{on } x = 0,\ L \\ u_0 = v_0 = w_0 = \beta_x = 0 & \text{on } y = 0,\ W \end{cases} \tag{49}$$

Immovable edge (SSSS3) :  $\quad u_0 = v_0 = w_0 = 0 \quad$  on all edges $\hfill (50)$

- Clamped support

$$\begin{cases} u_0 = v_0 = w_0 = \beta_x = \beta_y = 0 \\ w_{0,x} = w_{0,y} = 0 \end{cases} \quad \text{on all edges} \tag{51}$$

The Dirichlet boundary condition (BC) on $u_0, v_0, w_0, \beta_x$ and $\beta_y$ is easily treated as in the standard FEM, while the enforcement of Dirichlet BC for the derivatives $w_{0,x}, w_{0,y}$ can be solved as follows an idea of rotation-free of thin shell [48, 49]. The idea is to impose zero deflection for the control points, which are adjacent to the boundary control points.

Table 1 Material properties of functionally graded material

|  | $E$ (GPa) | $v$ | $k$ (W/mK) | $\alpha$ ($10^{-6}$ /K) | $\rho$ (kg/m$^3$) |
| --- | --- | --- | --- | --- | --- |
| Aluminium (Al) | 70 | 0.3 | 204 | 23 | 2707 |
| Alumina (Al$_2$O$_3$) | 380 | 0.3 | 10.4 | 7.2 | 3800 |
| Zirconia (ZrO$_2$) | 151 | 0.3 | 2.09 | 10 | 3000 |

For convenience, the following normalized transverse displacement, in-plane stresses and shear stresses are expressed as:

$$\bar{w} = \frac{w}{h}, \; \bar{\sigma} = \frac{\sigma h^2}{f_z a^2}, \; \bar{\tau} = \frac{\tau h}{f_z a}, \; \bar{P} = \frac{f_z a^4}{E_m h^4}$$

### *6.1 Nonlinear bending analysis*

In order to validate the present formulation, a moderate ($L/h = 10$) isotropic square plate ($v = 0.3$) subjected to a uniformly distributed load is first considered. Figure 5 shows the variation of the central deflection $\bar{w}$ versus load parameter $\bar{P}$ of this plate under two types of boundary conditions: SSSS1 and SSSS3. It can be seen that the present solutions are in good excellent with those of FEM reported by Reddy [41].

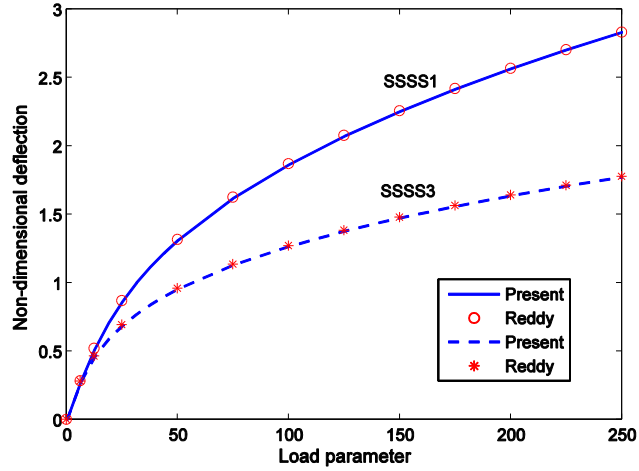

*Figure 5 The load-deflection curves of an isotropic square plate under SSSS1 and SSSS3 boundary conditions.*

Next, the geometrically nonlinear behaviour of Al/ZrO$_2$ plate in dimension as length $L$ = 0.2 m and thickness $h$ = 0.01 m is investigated. The plate is subjected to uniformly distributed load, which is increased sequential to equal to $f_z = -10^7$ N/m$^2$ after five steps. Figure 6 shows the variation of the load-central deflection curves via power index $n$. It should be noted that, index $n = 0$ corresponds to the ceramic plate, whilst $n = \infty$ indicates the metal plate. As expected, the deflection response of FGM plates is moderate for both linear and nonlinear cases compare to that of ceramic (stiffer) and metal (softer) plates. One more interesting point may be noted that the nonlinear deflections are smaller than linear ones and their discrepancy by increasing load. This is due to adding in the overall stiffer stiffness matrix by the nonlinear stiffness matrix $\mathbf{K}_{NL}$ which strongly depends on the deflection. Figure 7 plots the stress distributions through the plate thickness of the FGM plate ($n = 1$) via the change of load intensity. It can be seen that the effect of nonlinearity reduces the amplitude of the normalized stresses. Regarding the HSDT, the shear stress distributes as a curve with traction-free boundary condition at the top and bottom surfaces of the plate.

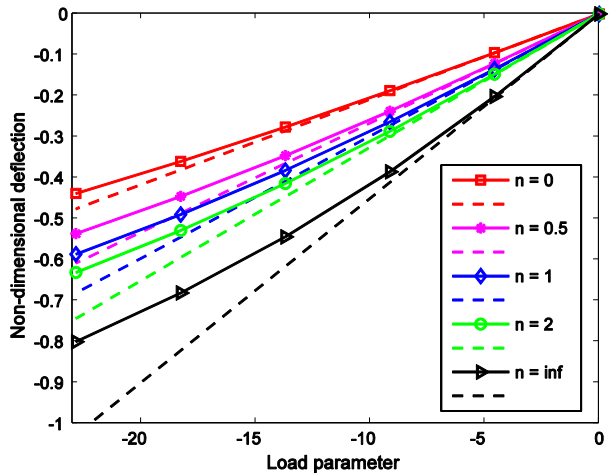

*Figure 6 Non-dimensional center deflection via load parameter and power index: nonlinear results (in solid line) and linear results (in dash line).*

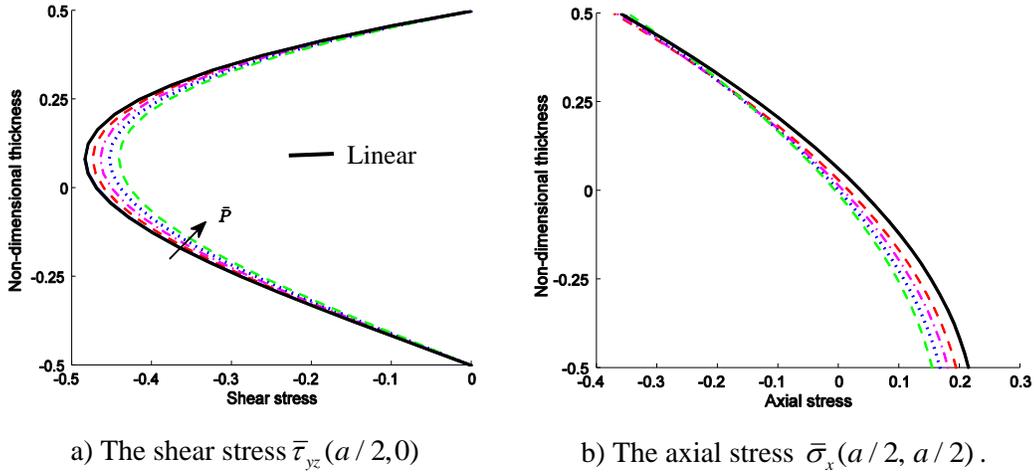

a) The shear stress $\bar{\tau}_{yz}(a/2,0)$

b) The axial stress $\bar{\sigma}_x(a/2, a/2)$.

*Figure 7 Effect of the load parameter $\bar{P}$ on the stresses distributions.*

By enforcing the temperature field to this plate as $T_m = 20°C$ and $T_c = 300°C$ at the bottom and top surfaces, respectively, the mechanical load – deflection curves via gradient index are plotted in Figure 8 in both cases of linear and nonlinear analyses. It is seen that the behaviour of deflection under thermo-mechanical load is quite different from purely mechanical load as shown in Figure 6. Because the higher temperature at the top surface causes the thermal expansion, the plates result in upward deflections. Among them, the metallic plate is found to be very sensitive to the temperature with the largest upward displacement. Then the deflection varies from positive side to negative side when the mechanical load increases. The similar tendency is observed for nonlinear analysis as compared with linear one except that the nonlinear deflections are larger than the linear ones under purely thermal load. This is due to the fact that development of the initial stress stiffness matrix $\mathbf{K}_0$, which is generated by thermal in-plane forces, reduces the overall plate stiffness. Another difference from linear solution is that the nonlinear results cannot be superimposed. For instant, as $n = 0$ the total deflection $\bar{w} = -0.3963$ is higher than a sum of $\bar{w} = -0.4385$ and $\bar{w} = 0.124$ in case of purely transverse and thermal load.

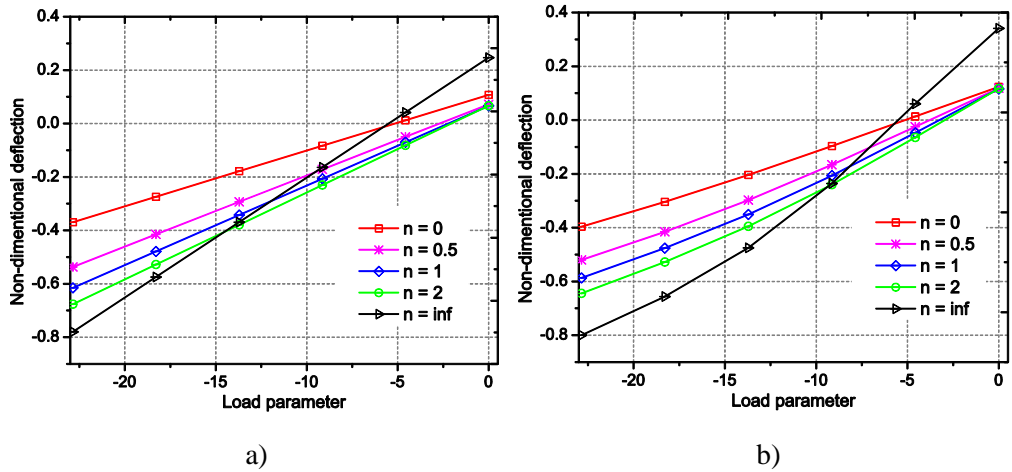

*Figure 8* Non-dimensional central deflection $\bar{w}$ of FGM plate under thermo-mechanical load via a) linear and b) nonlinear analyses.

Let's continuously investigate behaviour of the simply supported square Al/Al$_2$O$_3$ plate subjected to only thermal load. Figure 9 reveals the non-dimensional centre deflection via gradient temperature and power index. It can be seen that the plate is immediately bended toward the upper side as soon as temperature is enforced because of presence of extension-bending coupling effect due to un-symmetric material distribution through the thickness. For a comparison purpose, linear solutions are also supplied by neglecting the nonlinear stiffness matrix. It is observed that as temperature rises, increase in the thermal in-plane forces leads the plate stiffness to tend toward zero. As a result, the transverse displacement increases rapidly and runs to infinitive. It may be physically incorrect because the plate experiences large deflection at this time. So, von Karman nonlinear strain should be considered in the plate formulation. With the nonlinear effect, plate becomes stiffer and enables to bear higher temperature rise.

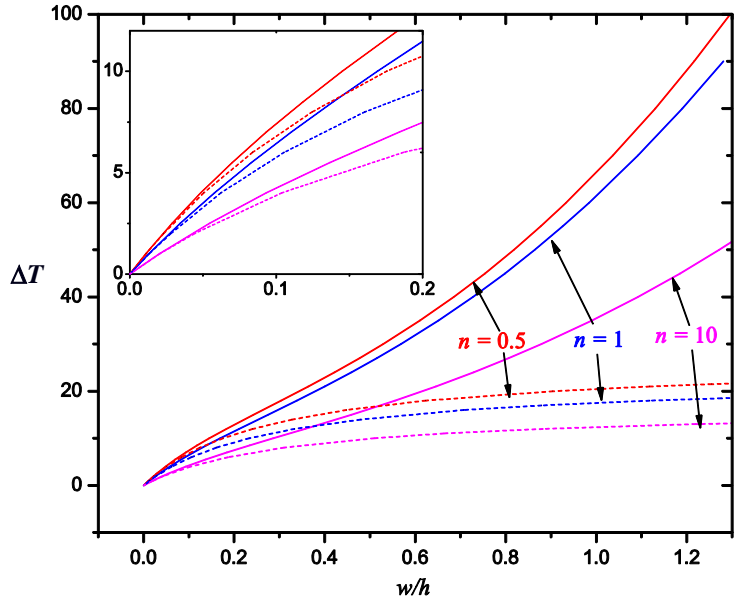

*Figure 9* Temperature-deflection curves of SSSS2 square Al/Al$_2$O$_3$ plate (L/h = 100) subjected to nonlinear temperature rise under linear analysis (in dash line) and nonlinear one (in solid line).

Figure 10 depicts the temperature – central deflection curves using various values of gradient index *n*. It is noted that homogeneous plates exhibit bifurcation buckling paths whilst FGM plates show no bifurcation phenomenon. Furthermore, decrease in the gradient index *n* increases the thermal carrying capability of the plate. In Figure 11, for comparison aim, the nonlinear bending behaviour of Al/Al$_2$O$_3$ plate (*n* = 1) under uniform, linear and nonlinear temperature rise is studied. Herein, the plate boundaries are constrained by two simply supported conditions: movable edges (SSSS1) and immovable edges (SSSS2). It is found that at an enough high temperature level, the uniform temperature distribution produces more transverse displacement in the plates than linear and nonlinear temperature distributions. In addition, movable edge condition (SSSS1) helps the plate to undergo smaller deformation than immovable edge one (SSSS2). Because weaker edge support and movability of in-plane displacements around all edges (except four corners), as shown in Figure 12, reduce the thermal effect on the plate. As noted that for clear vision, the in-plane displacements are scale by 1000.

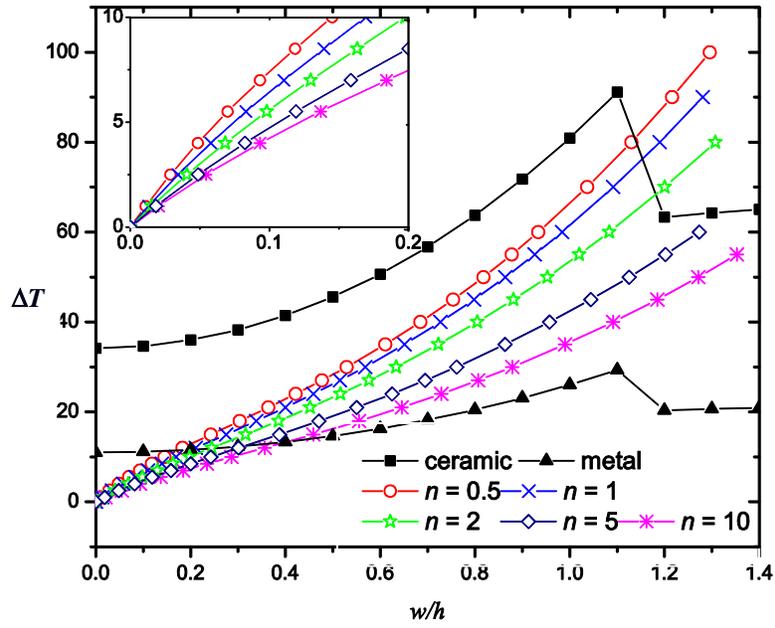

*Figure 10* Thermal post-buckling paths of SSSS2 square Al/Al$_2$O$_3$ plate (L/h = 100) under nonlinear temperature rise.

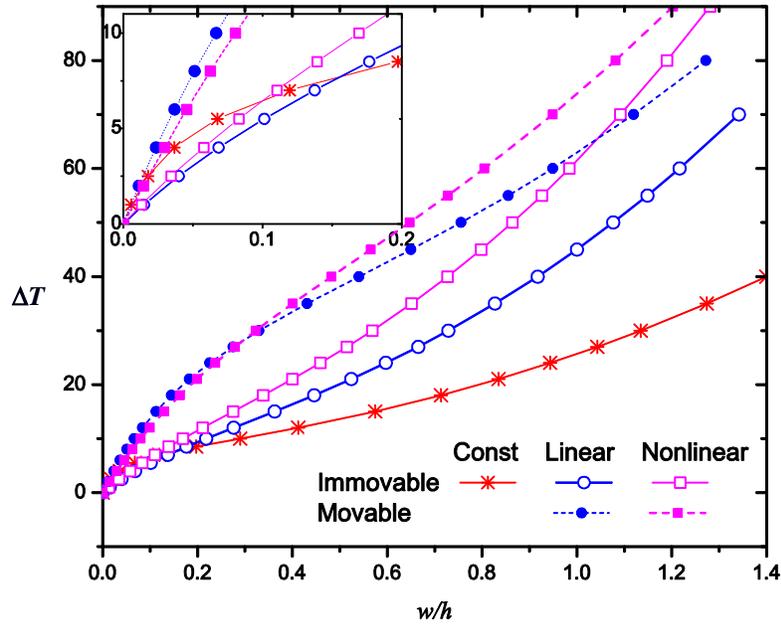

*Figure 11* Thermal post-buckling paths of the Al/Al$_2$O$_3$ plate (n = 1, L/h = 100).

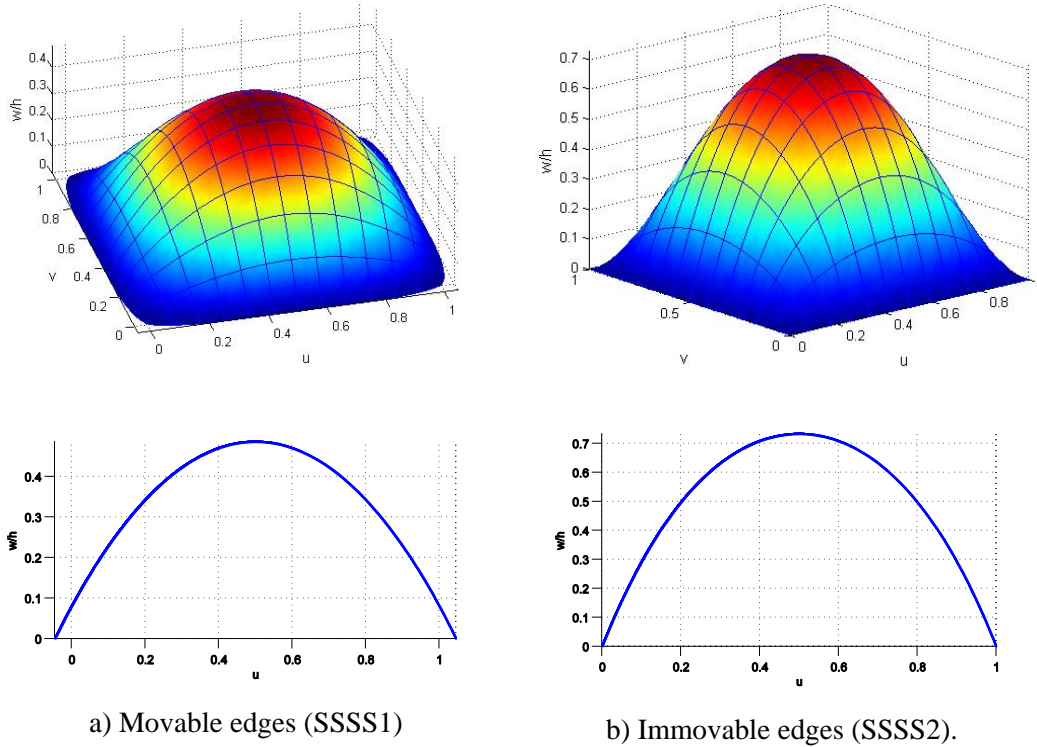

a) Movable edges (SSSS1)  b) Immovable edges (SSSS2).

*Figure 12 Displacement of Al/Al$_2$O$_3$ plate ( n = 1) at ΔT = 40°C under a) movable edge (SSSS1) and b) immovable edge (SSSS2) condition, whole plate profile (upper) and thermal deflection at cross section y = W/2 (lower).*

## *6.2    Thermal post-buckling analysis*

In this sub-section, two examples, for which solutions are available in the literature, are considered in order to validate the efficiency of the present method for the thermal instability. Firstly, the thermal post-buckling temperature − deflection curve of a simply supported square plate ($L/h = 10$, $v = 0.3$, $\alpha = 10^{-6}/°C$) under uniform temperature rise is plotted in Figure 13. The obtained results are compared with those of Bhimaraddi and Chandashekhara [50] using the parabolic shear deformation theory and the closed form solutions by Shen [24] based on higher-order shear deformation plate theory. Herein, it is evident that identical results are obtained in comparison with Shen's solutions for both perfect and imperfect plates (initial deflection $w^*/h = 0.1$). Herein, obtained critical temperature $\Delta T_{cr}^* = \alpha \Delta T_{cr} \times 10^4$ is as same as Shen's result [24] with the value of 119.783.

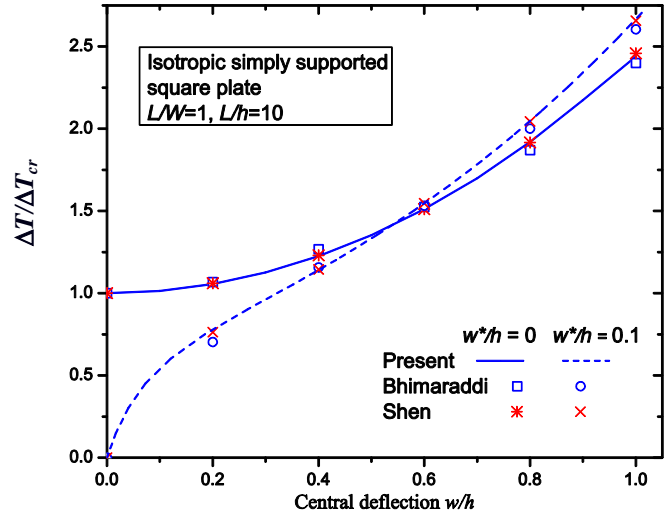

*Figure 13 Temperature-deflection curve of an isotropic square plate.*

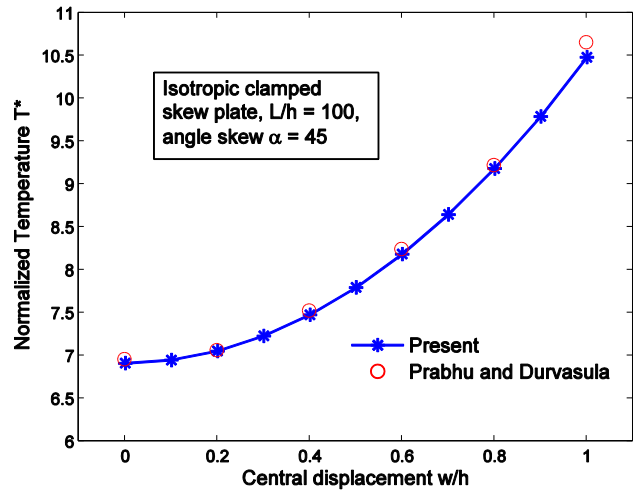

*Figure 14 Temperature-deflection curve of a clamped isotropic skew plate.*

Secondly, for the post-buckling path of a clamped skew plate (skew angle = 45°, $E$ = 1GPa, $\nu$ = 0.3, $\alpha$ = $10^{-6}/°C$) as depicted in Figure 14, the present solution is compared to that of Prabhu and Durvasula [51]. In this example, the temperature is normalized as $T^* = T_{cr} E \alpha L^2 h / (\pi^2 D)$ with the flexural rigidity $D = Eh^3 / 12(1-v^2)$. An excellent agreement is again observed.

Next, let us consider a clamped circular plate with radius-to-thickness ratio *R/h* = 100 subjected to uniform and nonlinear temperature rise. The plate is made from Al/Al$_2$O$_3$, for which material properties are assumed to be independent of temperature. The comparison of critical temperature of this plate is listed in Table 2. It is observed that the present results agree well with the closed-form solutions [52] and FEM's one [53] using a three-node shear flexible plate element based on the field-consistency principle as well as the solutions based on TSDT [35].

Table 2 Critical buckling temperature of FGM circular plate under temperature rise.

| *n* | Temp. Rise | Present | IGA [35] TSDT | FEM [53] FSDT | Closed form solution [52] FSDT | CPT |
|---|---|---|---|---|---|---|
| 0 | uniform | 12.7298 | 12.7247 | 12.713 | 12.712 | 12.716 |
|   | nonlinear | 25.4596 | 25.4494 | 25.426 | 25.924 | 25.433 |
| 0.5 | uniform | 7.2128 | 7.2107 | 7.203 | 7.202 | 7.204 |
|   | nonlinear | 19.0255 | 19.0193 | 18.996 | 18.996 | 19.002 |
| 1 | uniform | 5.9144 | 5.9128 | 5.907 | 5.906 | 5.907 |
|   | nonlinear | 15.3970 | 15.3929 | 15.377 | 15.373 | 15.378 |

Furthermore, Figure 15 shows the effect of power index *n* on the thermal post-buckling paths of the plates under the uniform and non-linear temperature rise. It should be noted that in case of nonlinear temperature rise, it is assumed that no temperature changes in the bottom of the plate, $\Delta T_m = 0$. Some following remarks are concluded:

- The thermal resistance of the FGM plates reduces due to increase in the material gradient index, *n*, because of the stiffness degradation by the higher metal inclusion, e.g., the thermal resistance is the highest if the plate is fully ceramic (*n* = 0) and the lowest if the homogeneous metal plate is retrieved (*n* = ∞).
- If we can keep the temperature varies non-uniformly through the thickness, FGM plates can resist higher buckling temperature.
- The clamped plates exhibit a bifurcation-type of instability, which is vertically symmetric.

- It is also observed that, after achieving the bifurcation point, the post-buckling temperature increases monotonically with the increase in the transverse displacement and suddenly drops to the secondary instability path. The transition from primary post-buckling path to the secondary one is caused by redistribution of post-buckling displacement mode shape. The maximum transverse displacement shifts from the plate centre towards one plate corner. This phenomenon can be seen in the reports for angle-ply composite plate by Singha et al. [54] and FGM plates by Prakash et al. [46, 47]. After the secondary instability, the post-buckling temperature slightly increases due to increase in deflection. This point is clearly illustrated in Figure 16.

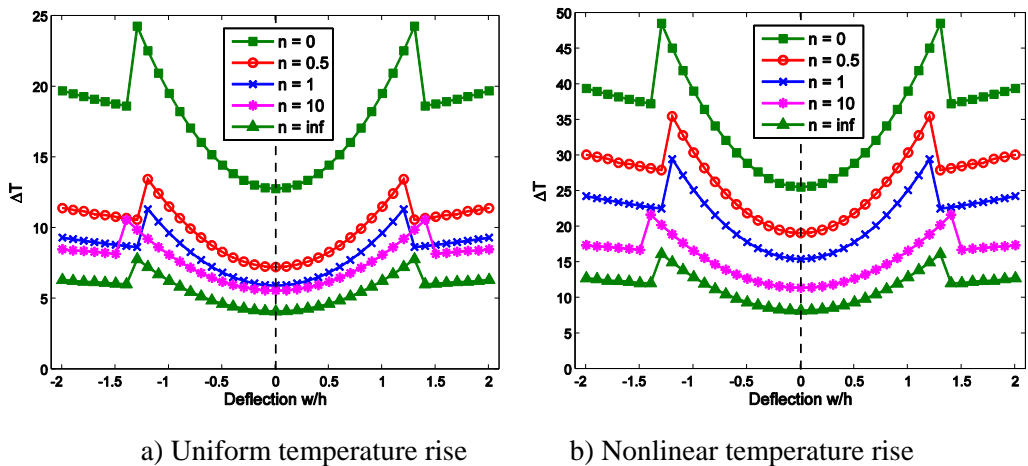

a) Uniform temperature rise  b) Nonlinear temperature rise

*Figure 15 Bifurcation buckling paths of the clamped circular Al/Al$_2$O$_3$ plate (R/h=100) under uniform and nonlinear temperature rise.*

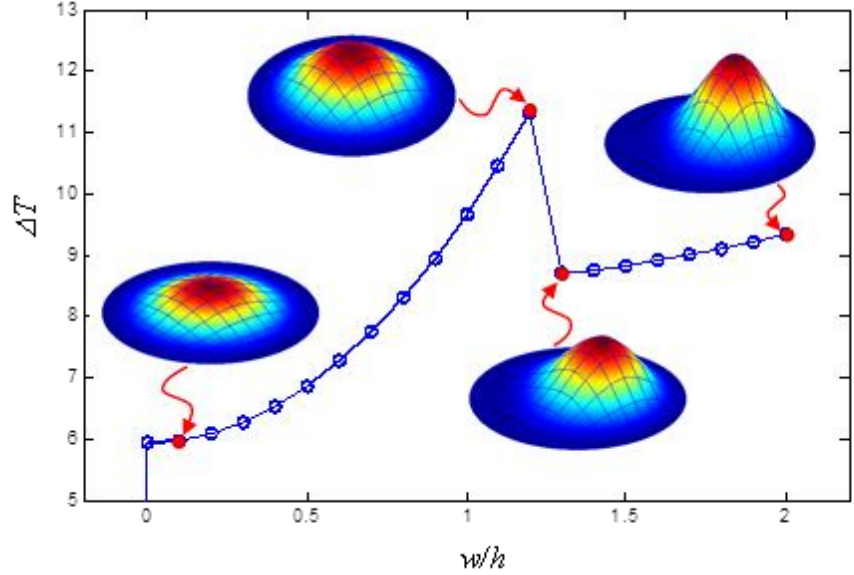

*Figure 16 Buckling modes of the clamped circular Al/Al$_2$O$_3$ plate (n =1, R/h = 100) under uniform temperature rise.*

### 6.3 Temperature-dependent material Si$_3$N$_4$/SUS304 plate

Finally, the thermal post-buckling of temperature-dependent material square plate, made of Silicon nitride (Si3N4) and Stainless steel (SUS304), is investigated. Their material properties are functions of temperature as indicated in Eq. (3) with the coefficients listed in Table 3 [39]. An example of the effect of temperature change on material properties of Si$_3$N$_4$/SUS304 FG plate, i.e. Young modulus is illustrated in Figure 17. It is observed that increase in temperature reduces Young modulus magnitude of both isotropic ($n = 0$) and FGM ($n = 1, 10$) plates.

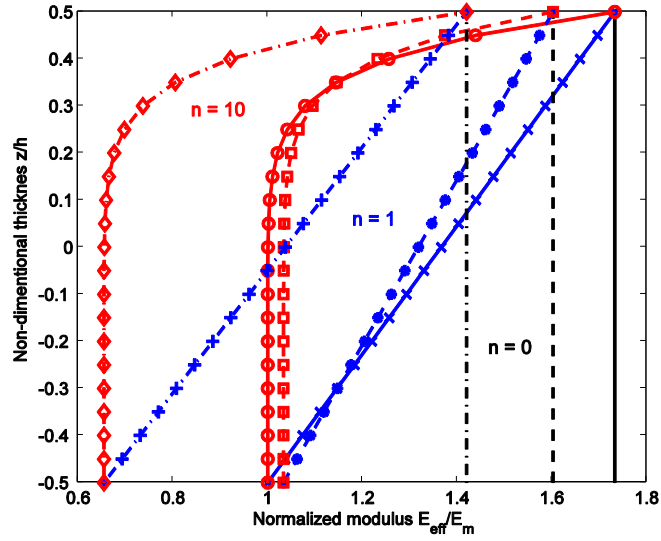

*Figure 17 The effective Young modulus of $Si_3N_4$/SUS304 plate at sepcified temperature: T=0 K (solid line), T=300 K (dashed line), T=1000 K (dash dot line).*

Table 3 Temperature dependent coefficients of $Si_3N_4$ and SUS304.

| Material | Property | $P_{-1}$ | $P_0$ | $P_1$ | $P_2$ | $P_3$ |
|---|---|---|---|---|---|---|
| Silicon nitride $Si_3N_4$ | $E$ (Pa) | 0 | 3.4843e11 | -3.0700e-4 | 2.1600e-7 | -8.946e-11 |
| | $\nu$ | 0 | 0.24 | 0 | 0 | 0 |
| | $\alpha$ (1/K) | 0 | 5.8723e-6 | 9.0950e-4 | 0 | 0 |
| | $k$ (W/mK) | 0 | 13.723 | -1.0320e-3 | 5.47e-7 | -7.88e-11 |
| Stainless steel SUS304 | $E$ (Pa) | 0 | 2.0104e11 | 3.0790e-4 | -6.534e-7 | 0 |
| | $\nu$ | 0 | 0.3262 | -2.00e-4 | 3.80e-7 | 0 |
| | $\alpha$ (1/K) | 0 | 1.2330e-5 | 8.0860e-4 | 0 | 0 |
| | $k$ (W/mK) | 0 | 15.379 | -1.26Ee-3 | 2.09e-6 | -7.22e-10 |

Figure 18 reveals the thermal post-buckling behaviours for $Si_3N_4$/SUS304 FGM plate with various power indices $n$ =0, 1, 10. The post-buckling paths for temperature-dependent and temperature-independent are presented in solid and dashed curves, respectively. Herein, the results considering temperature-independent material property (values are estimated at $T_0$ = 300K) are also presented for comparison purpose. It is observed that the thermal post-buckling curve becomes lower when considering the thermal dependent properties and increase in value of *n*. Furthermore, with thin plate (*L/h* = 100), the discrepancy between temperature-independent solutions and temperature-dependent solutions is insignificant due to the very small buckling temperature. As expected, with an increase in the length-to-thickness ratio, the critical buckling temperature increases accordingly.

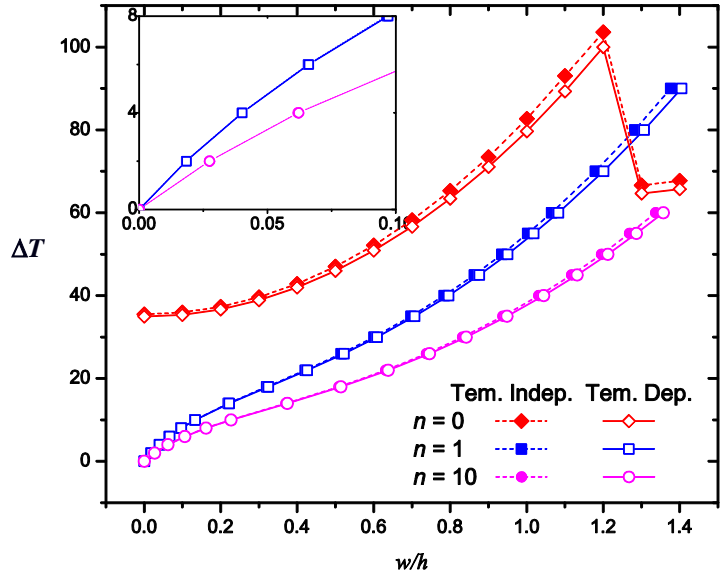

a) $L/h = 100$

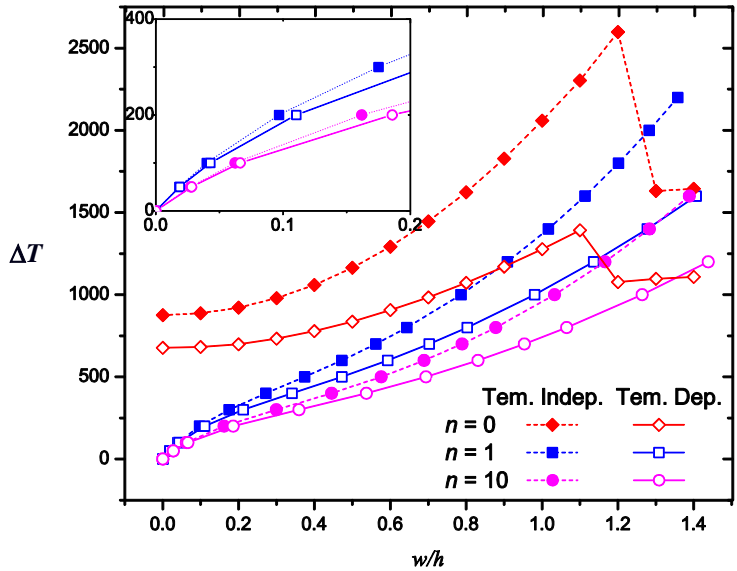

b) $L/h = 20$

*Figure 18* Thermal post-buckling paths of $Si_3N_4$/SUS304 FGM plate via various power indices and length-to-thickness ratios $L/h$.

## 7 Conclusion

This paper presents a simple and efficient formulation relied on the framework of NURBS-based IGA for nonlinear bending and post-buckling analysis of FGM plate in thermal environment. The material properties of the FGM plate are assumed to be the functions of both thickness position and temperature. The nonlinear governing equation of the plate is formed in the total Lagrange approach based on the von Karman assumptions. Due to value of force vector, this problem can be classified into two categories: geometrical nonlinear and nonlinear eigenvalue analyses. Through various numerical results, some concluding remarks can be drawn:

- There is a quite difference between linear and nonlinear solution. Under transverse load, nonlinear analysis achieves lower solutions because of additional nonlinear stiffness matrix. In case of purely thermal load, due to thermal membrane effect, the overall plate stiffness is reduced. As a result, the nonlinear deflections are larger than linear ones.

- In the FGM plate, temperature rise causes presence of the extension-bending effect due to its non-symmetric material properties. Therefore, no bifurcation type of instability occurs. However, in the special case, that is clamped boundary condition, the boundary constraint is capable to neutralize the extra moment. Thus, the buckling bifurcation does exist.

- The thermal resistance of the FGM plates reduces according to increase in the material gradient index *n* because of the stiffness degradation by the higher metal inclusion.

- FGM plate reduces the thermal resistance as temperature-dependent material properties are taken into account. This reduction is more clearly observed in thick plate.

**Acknowledgements**

The first author would like to acknowledge the support from Erasmus Mundus Action 2, Lotus Unlimited Project. The third author would like to acknowledge the support from National Research Foundation of Korea through grant NRF-2015R1A2A1A01007535.

**References**

[1] Miyamoto Y, Kaysser W, Rabin B, Kawasaki A, Ford R. Functionally graded materials: design, processing and applications: Springer Science & Business Media, 2013.

[2] Wang X, Dai H. Thermal buckling for local delamination near the surface of laminated cylindrical shells and delaminated growth. Journal of thermal stresses. 2003;26:423-42.

[3] Wang X, Lu G, Xiao D. Non-linear thermal buckling for local delamination near the surface of laminated cylindrical shell. International Journal of Mechanical Sciences. 2002;44:947-65.

[4] Koizumi M. FGM activities in Japan. Composites Part B: Engineering. 1997;28:1-4.

[5] Fukui Y, Yamanaka N. Elastic analysis for thick-walled tubes of functionally graded material subjected to internal pressure. JSME international journal Ser 1, Solid mechanics, strength of materials. 1992;35:379-85.

[6] Obata Y, Noda N. Transient thermal stresses in a plate of functionally gradient material Ceramic. Transactions of the Functionally Graded Materials. 1993;34:403-10.

[7] Praveen G, Reddy J. Nonlinear transient thermoelastic analysis of functionally graded ceramic-metal plates. International Journal of Solids and Structures. 1998;35:4457-76.

[8] Vel SS, Batra R. Exact solution for thermoelastic deformations of functionally graded thick rectangular plates. AIAA Journal. 2002;40:1421-33.

[9] Vel SS, Batra R. Three-dimensional exact solution for the vibration of functionally graded rectangular plates. Journal of Sound and Vibration. 2004;272:703-30.

[10] Javaheri R, Eslami M. Thermal buckling of functionally graded plates. AIAA Journal. 2002;40:162-9.

[11] Javaheri R, Eslami M. Thermal buckling of functionally graded plates based on higher order theory. Journal of thermal stresses. 2002;25:603-25.

[12] Ferreira A, Batra R, Roque C, Qian L, Jorge R. Natural frequencies of functionally graded plates by a meshless method. Composite Structures. 2006;75:593-600.

[13] Ferreira A, Batra R, Roque C, Qian L, Martins P. Static analysis of functionally graded plates using third-order shear deformation theory and a meshless method. Composite Structures. 2005;69:449-57.

[14] Park J-S, Kim J-H. Thermal postbuckling and vibration analyses of functionally graded plates. Journal of Sound and Vibration. 2006;289:77-93.

[15] Lee Y, Zhao X, Liew KM. Thermoelastic analysis of functionally graded plates using the element-free kp-Ritz method. Smart Materials and Structures. 2009;18:035007.

[16] Zhao X, Lee Y, Liew KM. Free vibration analysis of functionally graded plates using the element-free kp-Ritz method. Journal of Sound and Vibration. 2009;319:918-39.

[17] Nguyen-Xuan H, Tran LV, Nguyen-Thoi T, Vu-Do H. Analysis of functionally graded plates using an edge-based smoothed finite element method. Composite Structures. 2011;93:3019-39.

[18] Nguyen-Xuan H, Tran LV, Thai CH, Nguyen-Thoi T. Analysis of functionally graded plates by an efficient finite element method with node-based strain smoothing. Thin-Walled Structures. 2012;54:1-18.

[19] Phung-Van P, Nguyen-Thoi T, Tran LV, Nguyen-Xuan H. A cell-based smoothed discrete shear gap method (CS-DSG3) based on the C 0-type higher-order shear deformation theory for static and free vibration analyses of functionally graded plates. Computational Materials Science. 2013;79:857-72.

[20] Tran LV, Lee J, Nguyen-Van H, Nguyen-Xuan H, Wahab MA. Geometrically nonlinear isogeometric analysis of laminated composite plates based on higher-order shear deformation theory. International Journal of Non-Linear Mechanics. 2015;72:42-52.

[21] Reddy J. Analysis of functionally graded plates. International Journal for Numerical Methods in Engineering. 2000;47:663-84.

[22] Aliaga J, Reddy J. Nonlinear thermoelastic analysis of functionally graded plates using the third-order shear deformation theory. International Journal of Computational Engineering Science. 2004;5:753-79.

[23] Zaghloul S, Kennedy J. Nonlinear behavior of symmetrically laminated plates. Journal of Applied Mechanics. 1975;42:234-6.

[24] Shen H-S. Thermal postbuckling behavior of shear deformable FGM plates with temperature-dependent properties. International Journal of Mechanical Sciences. 2007;49:466-78.

[25] Liew K, Yang J, Kitipornchai S. Postbuckling of piezoelectric FGM plates subject to thermo-electro-mechanical loading. International Journal of Solids and Structures. 2003;40:3869-92.

[26] Nguyen-Xuan H, Tran LV, Thai CH, Kulasegaram S, Bordas SPA. Isogeometric analysis of functionally graded plates using a refined plate theory. Composites Part B: Engineering. 2014;64:222-34.

[27] Thai CH, Kulasegaram S, Tran LV, Nguyen-Xuan H. Generalized shear deformation theory for functionally graded isotropic and sandwich plates based on isogeometric approach. Computers & Structures. 2014;141:94-112.

[28] Hughes TJR, Cottrell JA, Bazilevs Y. Isogeometric analysis: CAD, finite elements, NURBS, exact geometry and mesh refinement. Computer Methods in Applied Mechanics and Engineering. 2005;194:4135-95.

[29] Cottrell JA, Hughes TJ, Bazilevs Y. Isogeometric analysis: toward integration of CAD and FEA: John Wiley & Sons, 2009.

[30] De Lorenzis L, Wriggers P, Hughes TJR. Isogeometric contact: a review. GAMM-Mitteilungen. 2014;37:85-123.

[31] Valizadeh N, Natarajan S, Gonzalez-Estrada OA, Rabczuk T, Bui TQ, Bordas SP. NURBS-based finite element analysis of functionally graded plates: static bending, vibration, buckling and flutter. Composite Structures. 2013;99:309-26.

[32] Yin S, Hale JS, Yu T, Bui TQ, Bordas SP. Isogeometric locking-free plate element: a simple first order shear deformation theory for functionally graded plates. Composite Structures. 2014;118:121-38.

[33] Tran LV, Ferreira AJM, Nguyen-Xuan H. Isogeometric analysis of functionally graded plates using higher-order shear deformation theory. Composites Part B: Engineering. 2013;51:368-83.

[34] Tran LV, Ly HA, Lee J, Wahab MA, Nguyen-Xuan H. Vibration analysis of cracked FGM plates using higher-order shear deformation theory and extended isogeometric approach. International Journal of Mechanical Sciences. 2015;96:65-78.

[35] Tran LV, Thai CH, Nguyen-Xuan H. An isogeometric finite element formulation for thermal buckling analysis of functionally graded plates. Finite Elements in Analysis and Design. 2013;73:65-76.

[36] Yin S, Yu T, Bui TQ, Nguyen MN. Geometrically nonlinear analysis of functionally graded plates using isogeometric analysis. Engineering Computations. 2015;32:519-58.

[37] Jari H, Atri H, Shojaee S. Nonlinear thermal analysis of functionally graded material plates using a NURBS based isogeometric approach. Composite Structures. 2015;119:333-45.

[38] Touloukian YS. Thermophysical Properties of High Temperature Solid Materials. Volume 4. Oxides and Their Solutions and Mixtures. Part I. Simple Oxygen Compounds and Their Mixtures. DTIC Document; 1966.

[39] Reddy J, Chin C. Thermomechanical analysis of functionally graded cylinders and plates. Journal of thermal stresses. 1998;21:593-626.

[40] Reddy JN. A Simple Higher-Order Theory for Laminated Composite Plates. Journal of Applied Mechanics. 1984;51:745-52.

[41] Reddy JN. Mechanics of laminated composite plates-theory and analysis. New York: CRC Press 2nd Edit, 2004.

[42] Najafizadeh M, Heydari H. Thermal buckling of functionally graded circular plates based on higher order shear deformation plate theory. European Journal of Mechanics-A/Solids. 2004;23:1085-100.

[43] Shariat BS, Eslami M. Thermal buckling of imperfect functionally graded plates. International Journal of Solids and Structures. 2006;43:4082-96.

[44] Turvey GJ. Buckling and postbuckling of composite plates: Springer Science & Business Media, 1995.

[45] Kiani Y, Bagherizadeh E, Eslami M. Thermal buckling of clamped thin rectangular FGM plates resting on Pasternak elastic foundation (Three approximate analytical solutions). ZAMM-Journal of Applied Mathematics and Mechanics/Zeitschrift für Angewandte Mathematik und Mechanik. 2011;91:581-93.

[46] Prakash T, Singha M, Ganapathi M. Thermal snapping of functionally graded materials plates. Materials & Design. 2009;30:4532-6.

[47] Prakash T, Singha M, Ganapathi M. Thermal postbuckling analysis of FGM skew plates. Engineering Structures. 2008;30:22-32.

[48] Benson D, Bazilevs Y, Hsu M-C, Hughes T. A large deformation, rotation-free, isogeometric shell. Computer Methods in Applied Mechanics and Engineering. 2011;200:1367-78.

[49] Nguyen-Thanh N, Kiendl J, Nguyen-Xuan H, Wüchner R, Bletzinger K, Bazilevs Y, et al. Rotation free isogeometric thin shell analysis using PHT-splines. Computer Methods in Applied Mechanics and Engineering. 2011;200:3410-24.

[50] Bhimaraddi A, Chandrashekhara K. Nonlinear vibrations of heated antisymmetric angle-ply laminated plates. International Journal of Solids and Structures. 1993;30:1255-68.

[51] Prabhu M, Durvasula S. Elastic stability of thermally stressed clamped-clamped skew plates. Journal of Applied Mechanics. 1974;41:820-1.

[52] Najafizadeh M, Hedayati B. Refined theory for thermoelastic stability of functionally graded circular plates. Journal of thermal stresses. 2004;27:857-80.

[53] Prakash T, Ganapathi M. Asymmetric flexural vibration and thermoelastic stability of FGM circular plates using finite element method. Composites Part B: Engineering. 2006;37:642-9.

[54] Singha MK, Ramachandra L, Bandyopadhyay J. Thermal postbuckling analysis of laminated composite plates. Composite Structures. 2001;54:453-8.